\documentclass[pre,twocolumn,amsfonts,amssymb,amsmath,letterpaper,nofootinbib,floatfix,showpacs]{revtex4}

\usepackage{graphicx}  
\begin{document}   
\newlength{\GraphicsWidth}
\setlength{\GraphicsWidth}{8cm}     

\newcommand\comment[1]{\textsc{{#1}}}

\renewcommand{\r}{\mathbf{r}}
\newcommand{\Man}{{\text{Manning}}}
\newcommand{\hr}{{\tilde{r}}}
\newcommand{\ha}{{\tilde{a}}}
\newcommand{\Imagin}{\Im m}
\newcommand{\eff}{{\text{eff}}}
\newcommand{\sat}{{\text{sat}}}
\newcommand{\be}{\begin{equation}}
\newcommand{\ee}{\end{equation}}

\title{%
Onsager-Manning-Oosawa condensation phenomenon and the effect of salt
}

\author{Emmanuel Trizac}
\affiliation{CTBP, 
UC San Diego, La Jolla (USA) and LPTMS Universit\'e Paris XI, 91405 Orsay (France)}
\author{Gabriel T\'ellez}
\affiliation{Departamento de F\'{\i}sica, Universidad de Los Andes,
A.A.~4976, Bogot\'a, Colombia}

\begin{abstract}                              
Making use of results pertaining to Painlev\'e III type 
equations, we revisit the celebrated Onsager-Manning-Oosawa condensation 
phenomenon for charged stiff linear polymers, 
in the mean-field approximation with salt. 
We obtain analytically
the associated critical line charge density, and show that it
is severely affected by finite salt effects, whereas previous results focused
on the no salt limit. In addition, we obtain explicit expressions
for the condensate thickness and the electric potential. The case of asymmetric
electrolytes is also briefly addressed.
\end{abstract}

\pacs{82.35.Rs, 61.20.-p, 87.15.-v}

\maketitle


Whereas scaling approaches have proven useful to describe neutral polymers, 
our understanding of polyelectrolytes solutions is quite rudimentary, due to
the long range character of Coulombic interactions \cite{Barrat,Levin}.
Stiff linear polyelectrolytes that are free of the coupling between 
chain conformation and small ions degrees of freedom consequently provide
ideal systems for a comprehensive comparison between theory and experiments.
There indeed exists a large variety of such rod-like polyions,
ranging from synthetic polymers (e.g those based on
poly(p-phenylene) backbones \cite{Holm1}) to biological molecules
(DNA, actin filaments, micro-tubules, some viruses etc). On distances smaller
than their (large) persistence length, these objects behave as charged cylinders
with an associated logarithmic electrostatic potential that may be strong
enough to bind oppositely charged microions (counterions). This was
first realized by Onsager and analyzed by Manning \cite{Manning} 
and Oosawa \cite{Oosawa}~: in the limit of vanishing polymer radius,
the corresponding phenomenon of counterion condensation is triggered when
the so-called Manning parameter $\xi= \ell_B/\ell$ is larger than unity
(see e.g. \cite{Barrat,Levin} for the essence of the argument).
Here, $\ell^{-1}$ is the backbone line charge in units of the elementary 
charge $e$ and $\ell_B = e^2/(\epsilon kT)$ denotes the Bjerrum length
with $\epsilon$ the dielectric constant of the solvent and $kT$ the
thermal energy.

Counterion condensation, which affects a gamut of static and dynamical
properties, is central to our view of highly charged polymers 
and an active field of research
\cite{Beyerlein,Rouzina,Tracy,Deserno,OS,Netz}. Interestingly, the 
influential arguments of Onsager, Manning and Oosawa found confirmation 
and a firm basis in numerical and analytical studies of the mean-field
non-linear Poisson-Boltzmann (PB) equation where the densities
of microions with valency $z$ are related to the local
mean electrostatic potential
$\varphi$ by $n_z \propto \exp(-z e \varphi /kT)$. The basic features are
already present in the exact solution of PB theory for a charged cylinder
in a concentric cylindrical Wigner-Seitz cell without salt \cite{Alfrey}, 
which can be cast 
as a restricted version of a partial differential equation first studied 
and solved by Liouville in 1853 \cite{Liouville}, as pointed out
in \cite{Fogden}. For an isolated cylinder (i.e. in the limit of 
a diverging Wigner-Seitz cell radius) with a small charge $\xi<1$, 
no counterions are bound to the cylinder and the potential distribution 
reduces to the bare logarithmic form. On the other hand, for 
$\xi>1$, a finite fraction of ions are ``bound'' to the surface
\cite{Netzapp}. 

The physically relevant situation where salt is present (i.e. with both co- and counter-ions,
that may come from the dissociation of an added electrolyte and/or from the
solvent itself) has resisted analytical 
understanding much longer. Ramanathan has showed that for $\xi>1$ and
distances much larger than the polyion radius $a$ the ionic atmosphere
was the same as that due to a cylinder with charge parameter 
$\xi=1$ \cite{Ramanathan}. 
This result holds asymptotically for $\kappa a \to 0$ \cite{Ramanathan,Fixman} 
where 
$\kappa^{-1}$ is the Debye length \cite{Barrat,Levin}. This 
condensation effect is the counterpart of the aforementioned 
one and was placed within the mathematical framework of isomonodromy 
theory in Ref. \cite{McCaskill} and Painlev\'e III equations 
in Refs. \cite{Davis,McCaskill,McCoy}.
This drew an exact correspondence with Ising model correlators
but more importantly allowed a) to compute the exact far-field
behaviour \cite{McCaskill}, b) to show rigorously the critical nature
of the value $\xi=1$ when $\kappa a \to 0$ \cite{McCaskill,Beyerlein}, 
and c) to obtain some analytical results for the potential distribution 
and ionic densities \cite{Tracy}. The common framework to the
previous approaches is PB theory \cite{rque} 
to which we will restrict ourselves
in the subsequent analysis. Most of existing results hold for 
$\kappa a \to 0$ only, and it appears that this (singular)
limit is approached logarithmically slowly. From a practical point
of view, finite $\kappa a$ corrections are therefore crucial and can never be
discarded. 

The purpose of the present work is to analyze the fate of the
counterion condensation phenomenon at finite salt concentration
($\kappa a\neq 0$). We shall show that for $\kappa a <1$, the transition 
is smoothed but remains, and that the associated critical charge 
parameter is salt-dependent and smaller than the usual Manning 
threshold $\xi_c=1$. The condensate structure will be resolved 
without the need to invoke any matching procedure, and analytically 
tractable results will be obtained for the electric potential 
below and above the critical value $\xi_c$. Extensive use will be made of
results pertaining to the theory of Painlev\'e III equations \cite{McCoy},
that mathematical difficulty have --to the best of our knowledge-- 
prevented to find their way towards the physicists' community
\cite{rque2}. The case of asymmetric 1:2 and 2:1 electrolytes
--that are physically not equivalent \cite{rque3}--
will also be briefly addressed.
Most of our results hold for 
$\kappa a<1$ and by comparison with numerical results can be shown to be
very accurate in this parameter range. 

We consider Poisson's equation in cylindrical coordinates
for the dimensionless electric potential
[$\phi = e \varphi/(kT)$]~:
\be
\frac{d^2 \phi}{d r^2} \,+\, \frac{1}{r} \frac{d \phi}{d r} \,=\, 
-\frac{\kappa^2}{z_+-z_-} \,\left[
e^{-z_+ \phi} - e^{-z_- \phi}
\right],
\label{eq:PB}
\ee
for the situations $(z_+,z_-)=(1,-1)$, $(1,-2)$ or $(2,-1)$, 
other asymmetries unfortunately resisting the analysis. 
We demand that $r d\phi/dr = -2 \xi$ 
for $r=a$ where the dimensionless bare charge of the polyion,
$\xi$, is assumed positive without loss of generality. 
The second boundary condition ($\phi \to 0$ when $r \to \infty$)
ensures that at large distances, $\phi$ obeys the linearized 
equation $\nabla^2 \phi = \kappa^2 \phi$ and therefore takes
the form 
\be
\phi(r) \, \underset{r\to\infty}{\sim} \,
\frac{2 \,\xi_{\eff}}{\kappa a \, K_1(\kappa a)} \, K_0(\kappa r),
\label{eq:farfield}
\ee
where $K_0$ and $K_1$ denote the modified Bessel functions 
of the second kind, of order 0 and 1 respectively. The prefactor
defining the {\it a priori} unknown effective charge $\xi_\eff$
reduces by definition to the bare charge $\xi$ for low values 
of $\xi$ (linear regime) but non-linear effects significantly
affect $\xi_\eff$. In particular, a consequence of counterion
condensation is that for $\kappa a \to 0$, the former quantity no
longer depends on $\xi$ provided $\xi>1$ 
\cite{Ramanathan,McCaskill,Beyerlein}.
We first focus onto 1:1 electrolytes for which the electric potential
$\phi_{11}$ crucially depends on a parameter $\lambda$ 
\cite{McCaskill,McCoy} 
that in the present context is related
to the effective charge through
\be
\xi_\eff \,=\, 2 \, \kappa a \,K_1(\kappa a)\, \lambda
\label{eq:xieff}
\ee
so that at large distances, $\phi \sim 4 \lambda K_0(\kappa r)$.
For $\lambda <1/\pi$ (which corresponds to $\xi<\xi_c$ where
$\xi_c$ is a critical charge to be defined below),
theorem 3 of Ref. \cite{McCoy} implies the short distance behaviour 
\be
e^{-\phi_{11}/2} \,=\, (\kappa r)^\sigma B \left[
1-\frac{(\kappa r)^{2-2\sigma}}{16 B^2 (1-\sigma)^2} 
\right] \,+\,
{\cal O}\left(\kappa r
\right)^2
\label{eq:petitpot}
\ee
where 
\be
B \,=\, 2^{-3 \sigma} \,\frac{\Gamma[(1-\sigma)/2]}{\Gamma[(1+\sigma)/2]},
\ee
$\Gamma$ being the Euler function. On the other hand, for $\lambda>1/\pi$, 
there is a qualitative change of behaviour for $\phi_{11}$ that then 
reads
\be
e^{-\phi_{11}/2} \,=\, \frac{\kappa r}{4 \mu}\, \sin\left[
2 \mu \log\left(\frac{\kappa r}{8}\right) - 2 \Psi(\mu)
\right] \,+\,
{\cal O}\left(\kappa r
\right)^4
\label{eq:grandpot}
\ee
where $\Psi$ denotes the argument of $\Gamma(i \mu)$ \cite{McCoy}.
For our purposes, a very accurate approximation 
--considered in the remainder--
is given by
the small $\mu$ Taylor expansion $\Psi(\mu) \simeq -\pi/2 - \gamma \mu$,
with $\gamma\simeq 0.5772...$ the Euler constant. Both quantities $\sigma$ 
(defined only for $\lambda<1/\pi$) and
$\mu$ (defined only for $\lambda>1/\pi$) 
are positive and related to the bare charge by the boundary condition 
$r \phi'(r)=-2\xi$ at $r=a$~: expressions (\ref{eq:petitpot}) 
and (\ref{eq:grandpot}) yield
\begin{eqnarray}
&&\xi \,=\, \sigma - \frac{(2-2\sigma) (\kappa a)^{2-2\sigma}}{16 (1-\sigma)^2
B^2-(\kappa a)^{2-2\sigma}} 
\label{eq:bc1}\\
&& (\xi-1) \tan\left[
2 \mu \log(\kappa a/8) +2 \mu \gamma 
\right] \,=\, 2 \mu
\label{eq:bc2}
\end{eqnarray}
where it is understood that $\mu$ is the smallest positive root
of (\ref{eq:bc2}).
As such, $\sigma$ and $\mu$ characterize the short distance features, 
and the difficulty amounts to connecting these parameters with
the far-field quantity $\lambda$. It may be shown that 
\cite{McCaskill,McCoy}
\begin{eqnarray}
&\displaystyle\lambda = \frac{1}{\pi} \, \sin\left(\frac{\pi \sigma}{2}\right) 
&\quad \hbox{for} \quad \lambda < \pi^{-1} ~(\hbox{ or }~ \xi<\xi_c)
\label{eq:connection1}\\
&\displaystyle\lambda = \frac{1}{\pi}\, \cosh(\pi \mu) 
&\quad \hbox{for} \quad \lambda > \pi^{-1} ~(\hbox{ or }~ \xi>\xi_c)
\label{eq:connection2}
\end{eqnarray}
The leading order term in $\phi_{11}$ is therefore $-2\sigma \log(\kappa r)$
for $\lambda<\pi^{-1}$ which corresponds to the bare potential of a cylinder
with line charge $\sigma$. For $\lambda>\pi^{-1}$ the dominant small
$r$ behaviour reads $-2 \log(\kappa r)$, up to an $r$-independent
term varying with charge and salt content \cite{rque4}. 

\begin{figure}[htb]
\includegraphics[height=4.4cm,angle=0]{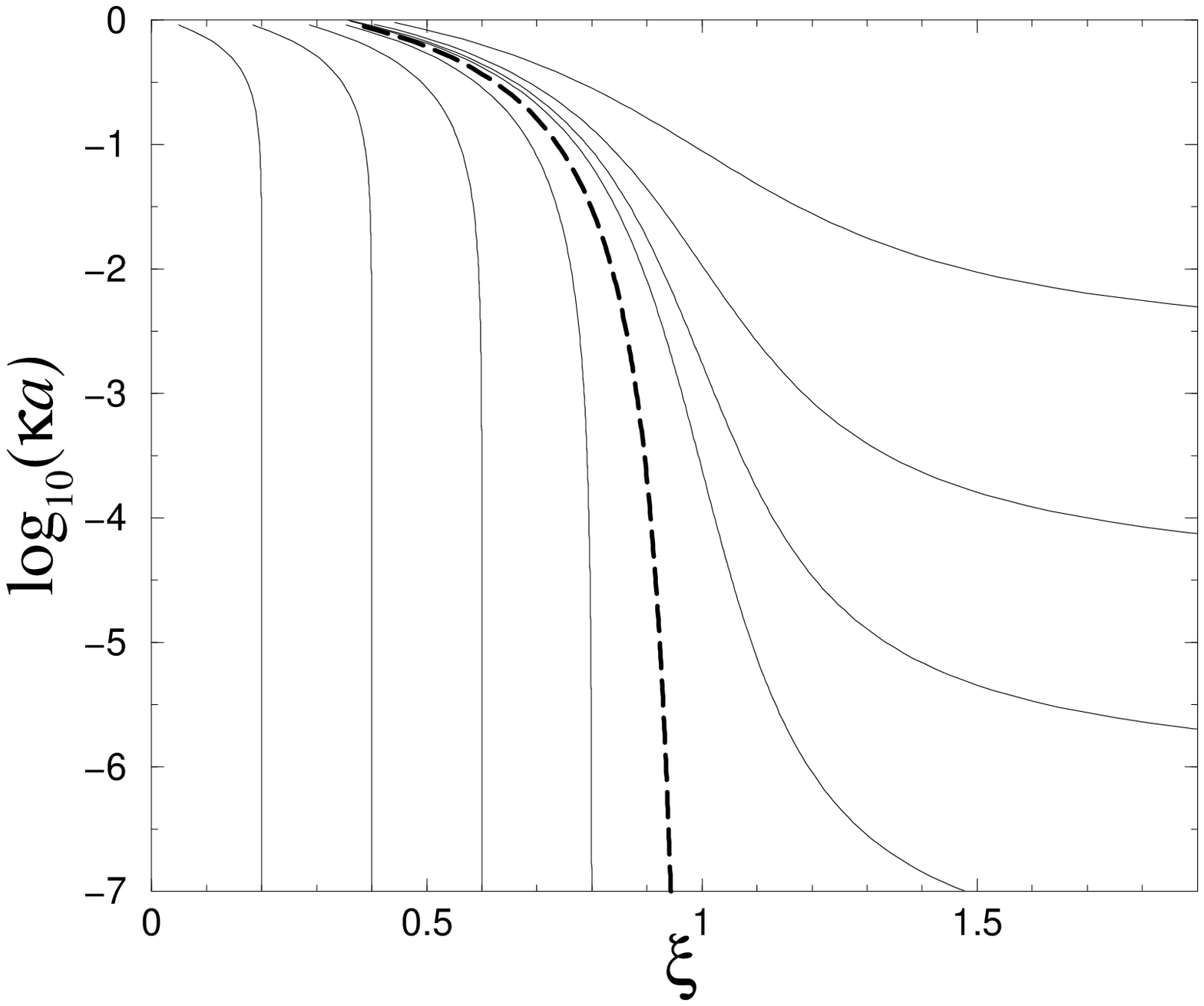}
\vskip -5mm
\caption{Contour lines for $\sigma$ and $\mu$ (and hence for $\lambda$) 
in the charge-salt plane,
as deduced from Eqs (\ref{eq:bc1}) and (\ref{eq:bc2}). 
The thick dashed curve shows the
locus of critical charges $\xi_c$ and separates the iso-$\sigma$ curves
for $\xi<\xi_c$ (from left
to right, the curves correspond to $\sigma=0.2,0.4,0.6$ and 0.8)
from the iso-$\mu$ curves shown --on the right hand side-- for $\xi>\xi_c$
(from bottom to top,  $\mu=0.08, 0.1, 0.13$ and 0.2).}
\label{fig:iso}
\end{figure}

\begin{figure}[hth]
\includegraphics[height=4.4cm,angle=0]{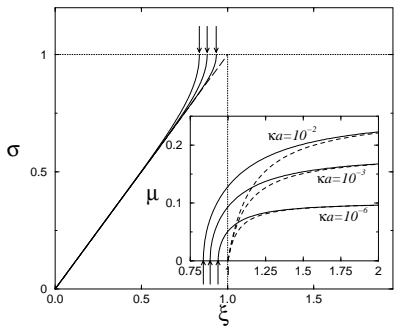}
\vskip -5mm
\caption{$\sigma$ versus bare charge. In the main graph and in the 
inset, 3 values of $\kappa a$ have been chosen. From left to right,
$\kappa a=10^{-2}, 10^{-3}$ and $10^{-6}$. For $\kappa a \to 0$,
$\sigma \to \xi$ (see the dashed first bissectrix) while $\mu \to 0$. 
The associated critical charges are indicated by the arrows (also in the
inset).
The inset shows $\mu$ versus $\xi$ 
and the dashed curves correspond to
approximation (\ref{eq:approxmu}). The vertical line indicates
$\xi=1$. }
\label{fig:sigmamudexi}
\end{figure}

The change of behaviour displayed by
Eqs. (\ref{eq:petitpot}) and (\ref{eq:grandpot}) is the fingerprint
of counterion condensation. The exponent $\sigma$ fulfills 
$0\leq\sigma\leq 1$ and
for the critical value $\lambda=1/\pi$, we have $\sigma=1$
while $\mu$ vanishes. The corresponding critical value of 
$\xi$ follows either from (\ref{eq:petitpot}) taking the limit 
$\sigma \to 1^-$ or enforcing $\mu\to 0^+$ in Eq. (\ref{eq:grandpot}).
These two routes yield exactly the same critical charge (which illustrates
the consistency of the underlying expressions for the potential):
\be
\xi_c \, =\, 1 \,+\, \frac{1}{\log(\kappa a)+\gamma -\log 8}.
\label{eq:xic}
\ee
In the limit $\kappa a \to 0$, the celebrated Manning threshold
$\xi_c=1$ is recovered but the correction embedded in 
(\ref{eq:xic}) is significant: at $\kappa a=10^{-3}$, $\xi_c$ is shifted 
to 0.881 and at
$\kappa a=0.1$, we get $\xi_c \simeq 0.737$. Expression (\ref{eq:xic})
has been derived omitting corrections of order $(\kappa r)^2$ in
(\ref{eq:petitpot}) and of order $(\kappa r)^4$ in
(\ref{eq:grandpot}). One may show however that the next correction 
to (\ref{eq:xic}) is of order $(\kappa a)^4 (\log \kappa a)^2$,
and is therefore irrelevant from a practical point of 
view whenever $\kappa a <1$ \cite{prep}. 
Figure \ref{fig:iso} may be considered as illustrating a 
``law of corresponding states'' and shows the values of bare charge
and salt concentration (or equivalently polyion radius) 
leading to the same electrostatic potential: if $\sigma$ or $\mu$ is fixed,
the connection formulae (\ref{eq:connection1}) and (\ref{eq:connection2})
indeed
ensure that $\lambda$ is also fixed, so that moving along the contour lines
of Fig \ref{fig:iso} leaves the full 
function $\phi(\kappa r)$ unaffected. The complementary information 
of the $\xi$ dependence of $\sigma$ and $\mu$ at fixed salt concentration
is shown in 
Figure \ref{fig:sigmamudexi}. It appears that except in the vicinity of
$\xi_c$, $\sigma$ is very close to $\xi$ (with systematically $\sigma>\xi$).
Close to the transition threshold, one has $\sigma-1$ and $\mu$ 
scaling like $|\xi-\xi_c|^{1/2}$. For highly charged polyions, 
the argument of the tangent function in (\ref{eq:bc2}) has to be close
to $-\pi$, from which we obtain
\be
\mu \,\simeq \, \frac{-\pi/2}{\log(\kappa a) +\gamma -\log 8 - (\xi-1)^{-1}}.
\label{eq:approxmu}
\ee
The inset of
Fig. \ref{fig:sigmamudexi} assesses the quality of this approximation
which improves when $\xi$ increases but fails quite severely when
$\xi < 1 + {\cal O}(1/\log \kappa a)$.

We have tested the accuracy of expressions 
(\ref{eq:petitpot})/(\ref{eq:bc1})/(\ref{eq:connection1})
below the critical charge, and Eqs. 
(\ref{eq:grandpot})/(\ref{eq:bc2})/(\ref{eq:connection2})
for $\xi>\xi_c$ against numerical solutions of the full non-linear 
PB equation (\ref{eq:PB}). These formulae turn out exceptionally accurate
for small $r$, while the far-field is conversely equally well given 
by (\ref{eq:farfield}) where $\xi_\eff$ is determined by 
(\ref{eq:xieff}). More 
interestingly, for $\kappa r <1/2$, $\phi_{11}$ deduced from 
(\ref{eq:petitpot}) is extremely accurate for all possible charges
and $\kappa a<0.1$ (curves not shown). 
Similarly, the far-field expression (\ref{eq:farfield})
provides very good accuracy as soon as $\kappa r>1/2$. It therefore
appears that $\phi_{11}$ is analytically known at all distances ; 
in the crossover region $\kappa r \simeq 0.5$ where no limiting
expression is supposed to be reliable, the worst relative accuracy
is observed but is nevertheless always below 2\%. 
We emphasize here that $\xi_\eff$ significantly differs from  
the critical charge $\xi_c$. In other words, even treating correctly 
the condensation phenomenon, non linear effects are still present
and affect the diffuse cloud of remaining ``uncondensed'' ions,
that cannot be treated within linearized PB or Debye-H\"uckel theory.
This important aspect has been overlooked so far
(see e.g. \cite{Manning}) and is present even in the simplest
limit $\kappa a \to 0$, where $\xi_\eff \to 2/\pi$ irrespective
of $\xi$ (provided $\xi>\xi_c$), whereas $\xi_c=1$.
The resulting repulsive force of interaction between two highly
charged rods is therefore overestimated by a factor ${\cal F}=(\pi/2)^2\simeq 2.5$
in the traditional Manning-Oosawa picture. 
An experimental check of this prediction has to 
fulfill the requirement of very low salt, since at say $\kappa a =10^{-2}$,
the above ratio is closer to unity [${\cal F} \simeq 1.4$ as can be checked from 
Eq. (\ref{eq:xieff})].

The previous results allow to discuss in a rigorous manner several
aspects of polyelectrolyte physics investigated in the literature.
Of particular interest is the so-called Manning radius $R_M$ 
\cite{Deserno,OS}, introduced to quantify the condensate thickness
when $\xi>\xi_c$. In the Wigner-Seitz cell approach, the integrated 
line charge per Bjerrum length $\ell_B$ [which, from
Gauss' law reads $q(r)=-r \phi'(r)/2$] has an inflexion point 
when plotted as a function of $\log r$ at the distance $R_M$ where
$q(R_M)=1$ \cite{Deserno,rque5}. It is remarkable that the $q(r)$ following
from (\ref{eq:grandpot}) shows the very same feature, which is ascribable
to a similar functional form, although at finite salt concentration
$q(a)\equiv\xi=1$ no longer provides the critical value of the charge.
The corresponding Manning radius reads:
\be
\kappa R_M \,=\, 8 \exp\left[-\pi/(4 \mu) -\gamma.
\right]
\label{eq:RMbon}
\ee
When $\xi>1+{\cal O}(1/\log \kappa a)$, one may use the approximation
(\ref{eq:approxmu}) and therefore
\be
\kappa R_M \,\simeq \, 2 \sqrt{2 \kappa a} \, \exp\left(
-\frac{\gamma}{2} -\frac{1}{2(\xi-1)}
\right).
\label{eq:bonbis}
\ee
The latter formula is the central result of a recent work \cite{OS}
where it was obtained by a clever but approximate matching procedure.
It appears to be incorrect when $\mu$ deviates from the large $\xi$
expansion (\ref{eq:approxmu}) (see the inset of Fig. \ref{fig:sigmamudexi}
where it is seen that the difference between the dashed and continuous curves
may be large). On the other hand, Eq. (\ref{eq:RMbon}) is always found to be 
very accurate compared to the numerical solutions of PB equation \cite{prep}. 
We note that choosing a different definition for the condensate thickness
$R^*$ (e.g. demanding that $q(R^*)=\xi_c$ instead of unity, leads 
for large bare charges to the scaling relation 
$R^* \propto a^\alpha \kappa^{\alpha-1}$ with now 
$\alpha=(\hbox{arctan}\,\pi)/\pi \simeq 0.402$,
smaller than the value $\alpha=1/2$ appearing in (\ref{eq:bonbis}).
The convenient inflection point criterion alluded to earlier would however
be lost following such a route.

We now turn to the case of asymmetric 1:2 and 2:1 electrolytes \cite{rque3}. 
Recent results obtained for a class of solutions
to the cylindrical Toda equations \cite{Toda} allow to 
extend the above analysis to such situations.
So far, the 1:2 case only has been studied, again only in the limit 
$\kappa a\to 0$ \cite{Tracy}. The details will be provided 
elsewhere \cite{prep};
we concentrate here on the critical charges
\begin{eqnarray} 	
\xi_c^{1:2} &=& \frac{1}{2} + \frac{1}{2 \log (\kappa a) + 2 \,{\cal C}^{1:2}}
\\
\xi_c^{2:1} &=& 1 + \frac{1}{\log (\kappa a) + {\cal C}^{2:1}}
\\
\hbox{with } {\cal C}^{1:2}\! &=&\! \gamma -(\log2)/3-3 (\log 3)/2 \simeq -1.301
\\
\hbox{and } {\cal C}^{2:1}\!&=&\!\gamma-\log2 -3 (\log 3)/2\simeq -1.763
.
\end{eqnarray}
The associated Manning radii, defined from 
$q(R_M) = 1$ (2:1 case) or $q(R_M) = 1/2$ (1:2 case) --which again ensures
the existence of an inflection point criterion-- read 
\begin{eqnarray}
\kappa R_M^{1:2} &\simeq& 2^{-1/3}\, 3^{3/4} \sqrt{2\kappa a} \, \exp\left[
-\frac{\gamma}{2} -\frac{1}{2(2\xi-1)}
\right]
\\
\kappa R_M^{2:1} &\simeq& 3^{3/4} \sqrt{2\kappa a} \, \exp\left[
-\frac{\gamma}{2} -\frac{1}{2(\xi-1)}
\right].
\end{eqnarray}
These expressions have the same status as (\ref{eq:bonbis}), i.e. 
hold at high enough $\xi$. It appears that the Manning 2:1 radius is inflated
a factor $3^{3/4}/2\simeq 1.14$ compared to its 1:1 counterpart,
irrespective of salt content and charge (but beyond the condensation threshold).
This quantifies the intuitive picture of a swollen double-layer
due to the presence of divalent coions (expelled further away
than monovalent ones), and 
conversely of a shrunk cloud 
(by a factor $2^{-4/3\,}3^{3/4}\simeq 0.90$ when $\xi$ is large enough)
in the 1:2
case due to the more efficient screening with divalent counterions. 

To conclude, exploiting an important body of mathematical work
in the field of stiff polyelectrolytes allows to systematically address
finite salt effects, that are crucial even under experimentally 
low salt conditions. The present work sheds new light on the
condensate structure 
and further connects short scale features with the long range
behaviour of the electric potential. Among the consequences of
experimental relevance following from our analysis, we mention
the large $q$ behaviour of the partial counterion/rod $S_-$
and coion/rod $S_+$ partial structure factors, expected to scale like
$q^{-2\pm 2\sigma}$ for $\xi<\xi_c$, with 
\smash{$S_-\stackrel{q\to\infty}{\propto} 1/\log q$}
at $\xi=\xi_c$. Such effects are
missed following the classical Manning-Oosawa arguments,
which generally result in an underestimation of screening.
\vskip -7mm



\begin{thebibliography}{99}
\bibitem{rque2}
PB theory may be cast into a Painlev\'e III problem 
for cylindrical polyions only \cite{McCaskill}.
\bibitem{rque3}
We denote by 1:2 the situation where coions are monovalent ($z_+=1$
for a positively charged polyion) and
counterions are divalent ($z_-=-2$ or more generally, $z_- =-2 z_+$). 
In the opposite 2:1 case, coions are divalent while counterions
are monovalent.
\bibitem{rque4}
We emphasize however that sub-leading corrections embodied in Eqs.
(\ref{eq:petitpot}) and (\ref{eq:grandpot}) are always relevant 
when numerical accuracy is pursued.
\bibitem{rque5}
The inflection point 
criterion may be extended beyond mean-field, see \cite{Deserno}.
\end{thebibliography}
\end{document}